\documentclass[reprint,twocolumn,superscriptaddress,preprintnumbers,amsmath,amssymb,aps,prd,floatfix]{revtex4-2}

\usepackage{diagbox}

\usepackage{mathtools}
\usepackage{amsfonts}
\usepackage{mathrsfs}
\usepackage{bbm}
\usepackage{slashed}
\usepackage{dsfont}

\usepackage{graphicx}
\usepackage{color}
\usepackage{array}

\usepackage{blindtext}
\usepackage{placeins}
\usepackage{booktabs}
\usepackage{makecell}
\usepackage{epstopdf}
\usepackage[caption=false]{subfig}
\usepackage{xspace}
\usepackage{siunitx}
\usepackage{hyperref}
\usepackage[nameinlink]{cleveref}
\usepackage{appendix}

\usepackage{xifthen}
\usepackage{xcolor}
\hypersetup{
	colorlinks,
	linkcolor={red!75!black},
	citecolor={blue!75!black},
	urlcolor={blue!75!black}
}

\setkeys{Gin}{width=0.48\textwidth}

\graphicspath{
	{./figures/}
}

\newcommand{\sqrtsNN}{\sqrt{s_{\small \textrm{NN}}}}
\newcommand{\sqrtsNNmath}{\sqrt{s_{\textrm{NN}}}}

\begin{document}

\title{Extracting freeze-out conditions in beam energy scan via functional QCD}

\author{Yi Lu}
%\email[]{qwertylou@pku.edu.cn}
\affiliation{Department of Physics and State Key Laboratory of Nuclear Physics and Technology, Peking University, Beijing 100871, China}

\author{Christian S. Fischer}
%\email[]{christian.fischer@theo.physik.uni-giessen.de}
\affiliation{Institut f{\"u}r Theoretische Physik, Justus-Liebig-Universit{\"a}t Gießen, 35392 Gießen, Germany}
\affiliation{
	Helmholtz Forschungsakademie Hessen f\"{u}r FAIR (HFHF),
	GSI Helmholtzzentrum f\"{u}r Schwerionenforschung,
	Campus Gie\ss{}en,
	35392 Gie\ss{}en,
	Germany
}

\author{Fei Gao}
%\email[]{fei.gao@bit.edu.cn}
\affiliation{School of Physics, Beijing Institute of Technology, 100081 Beijing, China}

\author{Yu-xin Liu}
%\email[]{yxliu@pku.edu.cn}
\affiliation{Department of Physics and State Key Laboratory of Nuclear Physics and Technology, Peking University, Beijing 100871, China}
\affiliation{Center for High Energy Physics, Peking University, 100871 Beijing, China}
\affiliation{Collaborative Innovation Center of Quantum Matter, Beijing 100871, China}

\author{Jan M. Pawlowski}
%\email[]{J.Pawlowski@thphys.uni-heidelberg.de}
\affiliation{Institut f{\"u}r Theoretische Physik,
	Universit{\"a}t Heidelberg, Philosophenweg 16,
	69120 Heidelberg, Germany
}

\affiliation{ExtreMe Matter Institute EMMI,
	GSI, Planckstr. 1,
	64291 Darmstadt, Germany
}

\begin{abstract}
Fluctuations of conserved charges provide a link between high quality theoretical results and precision measurements of 
heavy ion collisions. We compare results for ratios of the lowest order baryon number susceptibilities from functional 
QCD approaches to proton number cumulants extracted from experiments. We find that they meet at a specific temperature 
and chemical potential for each collision energy. This is indicative of the respective freeze-out point. From this 
self-consistent determination of the freeze-out parameters we extract a prediction for the kurtosis on the freeze-out line. 
We find quantitative agreement with experimental data where available, despite comparing apples (baryons) with oranges (protons). 
At a collision energy around 5 GeV, our kurtosis exhibits a peak structure indicative of the critical end point of QCD.
\end{abstract}

\maketitle

%%%%%%%%%%%%%%%%%%%%%%%%%%%%%%%
\textit{Introduction.---} 
%%%%%%%%%%%%%%%%%%%%%%%%%%%%%%%
%
The search for possible signatures for the critical end point (CEP) and the associated first order transition region in the phase 
diagram of quantum chromodynamics (QCD) is a primary goal of heavy ion collision experiments. For decades, great efforts have been 
made on measuring conserved charge fluctuations on an event by event basis, since these quantities have been suggested as probes to 
characterise the CEP of QCD, see e.g.~\cite{Luo:2017faz, Zhang:2026dny} for overviews. Recent high-precision extractions of 
net-proton number cumulants from beam energy scan (BES-II) experiments carried out by the STAR-collaboration at the Relativistic 
Heavy Ion  Collider (RHIC) \cite{STAR:2025zdq} together with older data from BES-I \cite{STAR:2021iop} emphasise the need for 
corresponding quantitative theoretical calculations from first principles. In our opinion, the way forward is to provide 
equilibrium QCD results for thermodynamical observables and fluctuations of conserved charges as a baseline, and then build upon 
this in non-equilibrium studies. The goal is to deliver a quantitative theoretical description of the experimental data that leads 
to a comprehensive understanding of the thermodynamic properties of QCD matter during the collisions. \\[-2ex]

%%%%%%%%%%%%%%%%%%%%%%%%%%%%%%%
\textit{Functional freeze-out curve.---} 
%%%%%%%%%%%%%%%%%%%%%%%%%%%%%%%
%
This programme requires as a first key milestone the determination of the freeze-out curve in the phase structure. 
The freeze-out condition has been widely investigated in statistical models through matching particle yields and their ratios, see e.g.~
\cite{Andronic:2017pug}. Alternatively, one can map out the freeze-out curve by comparing theoretical predictions for the ratios of susceptibilities of fluctuations of conserved charges with the experimental data. This strategy has been put forward at small chemical potential, comparing lattice results for ratios of net-baryon number fluctuations with the experimental data for ratios of net-proton number fluctuations, \cite{Bazavov:2012vg, Borsanyi:2013hza, Borsanyi:2014ewa, Bazavov:2015zja}. Note that such a comparison accommodates the differences between ratios of proton and baryon number fluctuations as well as the non-equilibrium part of the time evolution of heavy ion collisions into the map from \text{net-baryon} number fluctuations in \textit{equilibrium} to \textit{net-proton} number fluctuations at \textit{freeze-out}. 
This implicitly assumes that both effects approximately cancel out in ratios of fluctuations of conserved charges.   
Despite this caveat we consider this type of analysis an important stepping stone towards the complete picture including all effects. 

The functional approach to QCD offers the tantalising possibility to extend this approach to large baryon chemical potential. 
This includes the region of a potential critical end point (CEP) and/or the onset of new phases (ONP), such as the moat regime
\cite{Pisarski:2021qof, Fu:2024rto, Pawlowski:2025jpg} or inhomogeneous phases \cite{Buballa:2014tba, Motta:2024rvk, Pawlowski:2025jpg}. With full access to the first three
susceptibilities in the whole temperature--chemical-potential plane, a complete mapping between freeze-out parameters and 
collision energies can be built~\cite{Lu:2021ium}. In fact, one might argue that functional QCD methods are the only ones to date 
that allow for a direct computation of QCD thermodynamic quantities for large chemical potentials, see \cite{Lu:2025cls}. 
For $\mu_B/T \lesssim 4.5$, state of the art computations with the functional approach to QCD have quantitative 
reliability,~\cite{Fu:2019hdw, Gao:2020fbl, Gunkel:2021oya, Pawlowski:2025jpg}, and this border is continuously pushed towards 
larger chemical potentials. 

In the present work, we report on the extraction of a self-consistent \textit{functional} freeze-out condition in the context of STAR-data, 
which incorporates our current best knowledge on QCD thermodynamics at finite density. The resulting freeze-out curve allows us to estimate 
higher-order fluctuations of proton-number fluctuations with narrowed error bars, and in particular within the range of $3 < \sqrtsNN < 7.7$\,GeV.
To our knowledge, we deliver for the first time a prediction for the kurtosis at freeze-out in equilibrium QCD. This provides an important 
step towards isolating a smoking gun signal for the existence and the location of a QCD critical end point or more generally the onset of new phases. \\[-2ex]

%%%%%%%%%%%%%%%%%%%%%%%%%%%%%%%
\textit{Determination of the freeze-out parameters.---} 
%%%%%%%%%%%%%%%%%%%%%%%%%%%%%%%
%
The freeze-out temperature and chemical potential at the respective collision energy can be determined by comparing theoretical and 
experimental results on fluctuations of conserved charges. This requires the first two lowest-order ratios of baryon number susceptibilities
$\chi_2^B/\chi_1^B$ and $\chi_3^B/\chi_1^B$. They are compared with the corresponding cumulants of proton number that have been measured with 
relatively high statistics in BES-I~\cite{STAR:2021iop}, and with even higher precision in BES-II~\cite{STAR:2025zdq}. The freeze-out temperature $T_f$ and chemical potential $\mu_{B,f}$ at different collision energies $\sqrtsNN$ are determined by the 
following set of equations,
\begin{align}
  \frac{\chi_2^B}{\chi_1^B} \left( T_f,\mu_{B,f} \right) &= \frac{C_2}{C_1} \left( \sqrtsNN \right) \, , \label{eq:freezeout-extract-C1C2} \\[2ex]
  \frac{\chi_3^B}{\chi_1^B} \left( T_f,\mu_{B,f} \right) &= \frac{C_3}{C_1} \left( \sqrtsNN \right) \, , \label{eq:freezeout-extract-C3C1}
\end{align}
with the left hand side provided by theory and the right hand side by experiment. 
This method is inspired by previous studies in~\cite{Bazavov:2012vg, Borsanyi:2013hza, Lu:2021ium}, which offer a systematic and 
simultaneous extraction of the freeze-out temperature and baryon chemical potential. All technical details of this procedure are given 
in the Supplemental Materials. Here, we only discuss the key points of this method. 
On the one hand, it has been shown in lattice QCD calculations~\cite{Bazavov:2012vg,Borsanyi:2013hza} that the ratio $\chi_1^B/\chi_2^B$ is 
insensitive to temperature (even) in the vicinity of the chiral phase transition. Thus, it can be used to estimate the freeze-out chemical 
potential $\mu_{B,f}$ of high energy collisions. This is also verified in our calculations for the case of net-baryon charge, see the 
left panel of \Cref{fig:cumulant-ratio-compare-bes1} in the Supplemental Material for $\mu_B \in (0,150)\,$MeV. 
On the other hand, it is also understood~\cite{Lu:2021ium} that the ratio $\chi_3^B/\chi_1^B$ is more sensitive to temperature. This 
makes it a good candidate for estimating $T_f$, see the right panel of \Cref{fig:cumulant-ratio-compare-bes1}. The lattice QCD analysis
has been performed up to $\mu_B\sim150$\,MeV which includes the three data points at $\sqrtsNN = 200, 62.4, 39$ GeV~\cite{Borsanyi:2014ewa}. 
For large chemical potential, the equilibrium net-baryon number cumulants and their ratios are available from continuum QCD approaches via correlation functions, i.e. functional QCD~\cite{Fu:2023lcm, Lu:2025cls}. 
In this work, we adopt the approximation scheme developed in the latter work. It includes important aspects of confinement and chiral 
dynamics on a self-consistent basis and leads to quantitative agreement with lattice QCD benchmarks for thermodynamic quantities at 
small chemical potential.\\[-2ex]

\begin{figure}[t]
  \centering
  \includegraphics[width=0.95\columnwidth]{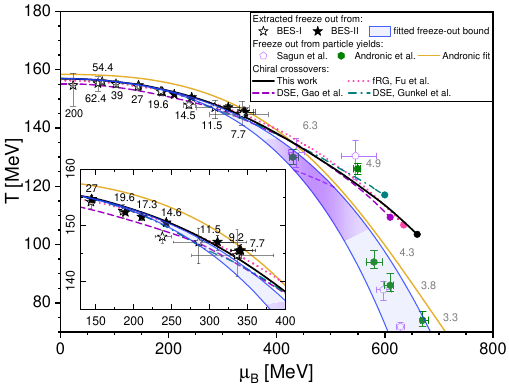}
  \caption{Freeze-out points extracted from \labelcref{eq:freezeout-extract-C1C2,eq:freezeout-extract-C3C1} in the $T$-$\mu_B$ plane, using the cumulant data $C_{1,2,3}$ from BES-I~\cite{STAR:2021iop} and BES-II~\cite{STAR:2025zdq}, with the collision energies labeled as the black numbers in GeV unit. Also shown are the locations of the CEP extracted in \cite{Fu:2019hdw, Gao:2020fbl, Gunkel:2021oya} and the corresponding result
  	\cite{Lu:2025cls} obtained with the scheme used in this work.
  	The prediction on the freeze-out condition towards lower $\sqrtsNN$ or higher $\mu_B$ is shown as the blue band, which is determined by fitting our freeze-out points with \labelcref{eq:muBfsNNfit,eq:TfmuBffit}. We also compare the freeze-out points with the chiral phase transition line, and the freeze-out points extracted from particle yields~\cite{Andronic:2017pug,Sagun:2017eye} whose collision energies are indicated by the grey numbers in GeV unit. The peak position of the kurtosis along the freeze-out lines within the band is indicated by the blue-dashed curve, with its peak width indicated by the purple region, in match with \Cref{fig:muBf-sqrtsNN} and \Cref{fig:kurtosis-along-extracted}.}
  \label{fig:freezeout-T-muB}
\end{figure}

%%%%%%%%%%%%%%%%%%%%%%%%%%%%%%%
\textit{An optimised estimate on the freeze-out line.---} 
%%%%%%%%%%%%%%%%%%%%%%%%%%%%%%%
%
We compare the susceptibility ratios $\chi_2^B/\chi_1^B$ and $\chi_3^B/\chi_1^B$  with the data from both BES-I and BES-II 
experiments. As depicted  in \Cref{fig:freezeout-T-muB}, we find that for high and intermediate $\sqrtsNN \gtrsim 10\,$GeV, the 
extracted freeze-out points agree with the chiral crossover line within error bars. Details on the latter are 
explained in the Supplemental Material. We take into account both the statistical and systematic errors of the experimental data. Furthermore, we take into account errors from the extraction scheme \labelcref{eq:freezeout-extract-C1C2,eq:freezeout-extract-C3C1}. For smaller collision energies, our combined error bars become larger.    

The consistency between the extracted freeze-out points and the phase transition line at small chemical potentials suggests that one can parametrise the freeze-out line similar to the chiral phase transition line. This reads~\cite{Cleymans:2005xv}, 
\begin{align}
T_{f} = T_0 \left[ 1 - \kappa_{2}^{f} \left(\frac{\mu_{B,f}}{T_0}\right)^2 - \kappa_{4}^{f} \left(\frac{\mu_{B,f}}{T_0}\right)^4 + \cdots \right]\,,
\label{eq:TfmuBffit}
\end{align}
with $T_0$ the freeze-out temperature at zero chemical potential. The curvature coefficients $\kappa_{2}^{f}, \kappa_{4}^{f}, \cdots$ accommodate the $\mu_{B,f}/T_0$-dependence of the freeze-out temperature. These parameters of the freeze-out line are in fact strongly constrained by mainly two aspects:
\begin{itemize}
\item[\emph{(i)}] The direct extraction results in a freeze-out line which is consistent with the chiral phase transition line. It exhibits a  milder $\mu_B$ dependence though, yielding the relations $T_0=T_c^\chi(\mu_B=0)$ and $\kappa^f_2 \leq \kappa_2^\chi$ with 
the corresponding parameters $T_c^\chi$ and $\kappa_2^\chi$ of the chiral phase transition line.
\item[\emph{(ii)}] It is a natural physical assumption that the chemical freeze-out takes place after the relatively sharp chiral crossover. In the vicinity of small $\mu_B$ and with $T_0 = T_c^\chi(0)$, this requires $\kappa^f_2 \geq \kappa_2^\chi$. 
\end{itemize}
The combination of the two constraints entails that the two parameters of the freeze-out line are tightly constrained:
\begin{align}
T_0=T_c^\chi(0)=157\,\rm{MeV};\quad \kappa^f_2 =\kappa_2^\chi=0.0153,
\end{align}
with small negligible errors.
In turn, the extrapolation of the fit beyond $\mu_B \geq 350\,$MeV is not well constrained by our freeze-out data even with the physics constraint mentioned above. Consequently, the higher-order terms in the $\mu_B$ expansion \labelcref{eq:TfmuBffit} are less constrained. 
A minimal parametrisation in the high $\mu_B$ region is given by one additional expansion parameter, $\kappa_{4}^{f}$. Its deviation from 
$\kappa_{4}^{\chi} \approx 0$ \cite{HotQCD:2018pds, Borsanyi:2020fev} determines the deviation between the freeze-out and the phase transition
line. 
It will be fixed mainly by the freeze-out points at lower collision energies. We take into account the 6.3\,GeV freeze-out point, extracted from particle yields in the SPS experiment~\cite{Andronic:2017pug}, which is discussed as a reliable reference in the 
intermediate $\mu_B$ region, together with our freeze-out points from BES-I 200 to 27 GeV and BES-II 27 to 7.7 GeV in
\Cref{fig:freezeout-T-muB}. 

\begin{figure}[tb]
  \centering
  \includegraphics[width=0.95\columnwidth]{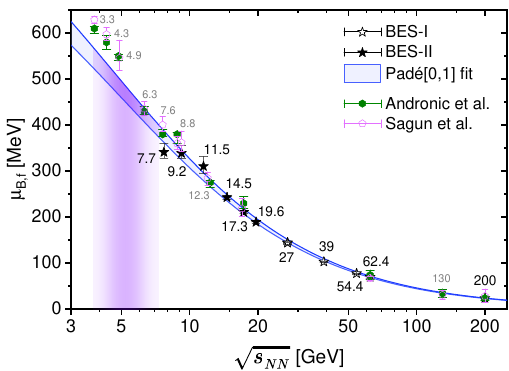}
  \caption{Freeze-out chemical potential $\mu_{B,f}^{}$ as a function of collision energy $\sqrtsNN$. 
  Our result for the $\mu_{B,f}^{}$-$\sqrtsNN$ relation is displayed as the blue band, which is compared with the freeze-out points predicted from statistical models~\cite{Andronic:2017pug,Sagun:2017eye}. The collision energies in STAR BES-I and BES-II are marked in black, and those collision energies covered in other experiments are marked in gray, all in GeV units. }
  \label{fig:muBf-sqrtsNN}
\end{figure}

To connect our results with the experimental landscape in terms of collision energy, we also exploit knowledge on the $\mu_{B,f}$\,-\,$\sqrtsNN$-dependence. A commonly used fit is the Pad\'{e}[0,1] fit~\cite{Andronic:2017pug,Cleymans:2005xv}, %
\begin{align}
\mu_{B,f} = \frac{a}{1+b \sqrtsNNmath }\,,
\label{eq:muBfsNNfit}
\end{align}
which describes our extracted freeze-out data well as depicted in \Cref{fig:muBf-sqrtsNN}.

\begin{table}[tb]
\setlength{\tabcolsep}{4pt}
\renewcommand{\arraystretch}{1.4}
\begin{tabular}{c|c|c|c|c|c}
\hline\hline
      & $a$ {[}MeV{]} & $b$ {[}GeV$^{-1}${]} & $T_0$ {[}MeV{]} & ~$\kappa_{2}^{f}$~     & $\kappa_{4}^{f}$     \\
\hline
lower & 913.9         & 0.1977               & $T_c^\chi(0)$   &   ~$\kappa_2^\chi$~    &   1.47$\times 10^{-3}$   \\
\hline
upper & 1027.0        & 0.2143               & $T_c^\chi(0)$   &   ~$\kappa_2^\chi$~    &   0.73$\times 10^{-3}$   \\
\hline
mean & 953.7        & 0.2026               & $T_c^\chi(0)$   &   ~$\kappa_2^\chi$~    &   0.96$\times 10^{-3}$   \\
\hline\hline
\end{tabular}
\caption{Fit parameters for the upper and lower freeze-out bounds and the best fit from mean freeze-out data. $T_c^\chi(0) = 157\,$MeV and $\kappa_2^\chi = 0.0153$ are the pseudo-critical temperature at zero $\mu_B$ and the curvature of the chiral crossover line, respectively, following the functional QCD calculation in~\cite{Lu:2025cls}.}
\label{tab:freezeout-fit}
\end{table}

The error bars of the freeze-out points are included in the fit of \labelcref{eq:muBfsNNfit,eq:TfmuBffit}, which provide upper and lower 
bounds on the freeze-out parameters. Specifically, the upper/lower bound is determined by fitting the upper/lower bounds of both $T_{f}$ 
and $\mu_{B,f}$, respectively. 
The results are collected in \Cref{tab:freezeout-fit}, and the upper and lower bounds of the freeze-out line are visualised as the blue 
bands in \Cref{fig:freezeout-T-muB} for $T_f(\mu_{B,f}^{})$ and in \Cref{fig:muBf-sqrtsNN} for $\mu_{B,f}^{}(\sqrtsNN)$. 
Our results confirm the validity of the $T_f(\mu_{B,f}^{})$ freeze-out extrapolation for AGS collision energies down to 
$\sqrtsNN \approx 3\,$GeV: our freeze-out bound covers precisely the freeze-out points predicted by the particle yields from the 
statistical models~\cite{Andronic:2017pug,Sagun:2017eye}, the latter marked out in \Cref{fig:freezeout-T-muB} with the collision 
energies by the grey numbers in units of GeV. In particular, the upper bound is already close to the freeze-out line from a global 
fit analysis in~\cite{Andronic:2017pug}. 

Our $\mu_{B,f}(\sqrtsNN)$-curve also agrees well with the predictions from statistical models for $\sqrtsNN \leq 7.7\,$GeV. 
Note that the $\mu_{B,f}(\sqrtsNN)$ error band is correlated with the $T_f(\mu_{B,f}^{})$ band, as both their upper/lower 
bounds reflect the minimally/maximally possible freeze-out chemical potential at a given collision energy. Considering our extrapolation 
down to even lower collision energies $\sqrtsNN \lesssim 5\,$GeV, we find a slightly lower $\mu_{B,f}$ compared to those of 
statistical models. 
This discrepancy is interesting and warrants further studies, both on the experimental and theoretical side. Even taking into account
the large error bars of the freeze-out predictions from different statistical models, it might indicate that 5\,GeV marks the onset of new dynamics. 
\\[-2ex]

\begin{figure}[t]
  \centering
  \includegraphics[width=0.95\columnwidth]{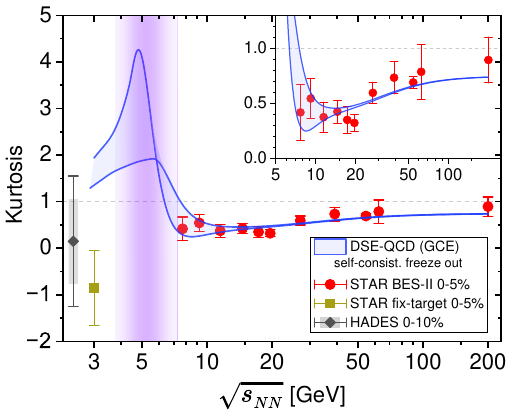}
  \caption{Collision energy ($\sqrtsNN$) dependence of the net-baryon number kurtosis along the fitted freeze-out band in \Cref{fig:freezeout-T-muB}. The error estimate of the kurtosis is marked by the blue band, which corresponds to its variation along all possible freeze-out lines within the freeze-out band in \Cref{fig:freezeout-T-muB}. }
  \label{fig:kurtosis-along-extracted}
\end{figure}
%
%%%%%%%%%%%%%%%%%%%%%%%%%%%%%%%
\textit{Prediction for the kurtosis.---} 
%%%%%%%%%%%%%%%%%%%%%%%%%%%%%%%
%
The freeze-out estimate further allows an improved prediction of the kurtosis baseline, which is shown in
\Cref{fig:kurtosis-along-extracted}. The $\sqrtsNN$-dependence of the kurtosis along the upper/lower bound of the freeze-out band 
is marked by the solid blue line, and the kurtosis along the upper bound has a higher peak compared to that along the lower ones. 
We also consider a linear interpolation between the upper and lower fit parameters in \Cref{tab:freezeout-fit}, which defines a 
continuous change of the freeze-out line within the band. The error estimate of the kurtosis corresponds to its variation along 
all possible freeze-out lines in between. This is shown as the blue band.

The agreement on the experimental data further supports that chemical freeze-out occurs close to the chiral phase transition for $\sqrtsNN \gtrsim 10\,$GeV. 
For $\sqrtsNN$ from 7.7 down to 3\,GeV, our improved freeze-out analysis predicts a peak of the kurtosis at $\sqrtsNN\approx$ 5 to 6\,GeV.
This location agrees with previous estimates from functional QCD~\cite{Fu:2023lcm,Lu:2025cls}, however the peak height is lower. 
The freeze-out temperature and chemical potential at the peak of kurtosis are also extracted within the upper and lower bounds 
given by \Cref{tab:freezeout-fit}, which are displayed as the purple dashed curve in \Cref{fig:freezeout-T-muB}. In particular, we find 
\begin{align}
 (T_f,\mu_{B,f}) = (120,506)\,\mbox{MeV}\,,\quad  \sqrtsNN  = 5\,\mbox{GeV}
\end{align} 
for the upper bound, and
\begin{align}
 (T_f,\mu_{B,f}) = (125,432)\,\mbox{MeV}\,,\quad  \sqrtsNN = 6\,\mbox{GeV} 
\end{align}
for the lower bound. 
We also provide an estimate on the width of the kurtosis peak, which is evaluated from a Breit-Wigner fit along the upper or lower 
bound of freeze-out. The overall width spread in the $\sqrtsNN$-direction is shown as the purple area in \Cref{fig:kurtosis-along-extracted}. 
The kurtosis peak yields a potential ``smoking-gun'' area in the experimental landscape, and its projection in the $(T,\mu_B)$ plane is marked out in \Cref{fig:freezeout-T-muB} as the purple region. 
Notably, this region is also located at the onset position $\mu_B \gtrsim 450$\,MeV where the freeze-out bound starts to drop much quicker than the chiral crossover line toward larger chemical potential. \\[-2ex]

%%%%%%%%%%%%%%%%%%%%%%%%%%%%%%%
\textit{Summary.---} 
%%%%%%%%%%%%%%%%%%%%%%%%%%%%%%%
%
The current situation regarding the QCD phase diagram and the potential identification of a critical end point is complex.
On the theory side, the estimates for the location of a possible critical end point or the onset of new phase from functional QCD, 
\cite{Fu:2019hdw, Gao:2020fbl, Gunkel:2021oya, Pawlowski:2025jpg}, or model computations such as in holographic models for QCD \cite{Hippert:2023bel} and extrapolations of the 
positions of Lee--Yang zeros and thermodynamic quantities in lattice QCD \cite{Basar:2023nkp, Clarke:2024ugt, Shah:2024img}, converge rather impressively. 
On the experimental side, recent BESII fluctuation data \cite{STAR:2025zdq} reveal no critical information. In this work we
demonstrate that both findings are not in disagreement with each other. 

Systematically comparing experimental data on ratios of net-proton cumulants with results for rations of baryon susceptibilities 
from functional methods, we have mapped out the freeze-out curve in the QCD phase diagram. We find consistent overlap with extractions
using statistical models for collision energies above $\sqrtsNN \ge 7$ GeV, i.e. in the `base-line' region of the kurtosis. 
We also find interesting discrepancies at small collision energies $\sqrtsNN \le 5$ GeV. They warrant further investigation, since this might signal the onset of new phases such as a moat regime and/or regions with spatial inhomogeneities not yet covered by our study. 
For temperatures in the vicinity of our CEP, we find a distinct deviation of the freeze-out line from the chiral transition line 
as envisaged e.g.~in Ref.~\cite{Stephanov:2008qz}. Moreover, we find a peak in the kurtosis in the region 
$(T_f,\mu_{B,f}) = (120,506)\,$MeV, $\sqrtsNN  = 5\,$GeV for the upper bound of our error bar and 
$(T_f,\mu_{B,f}) = (125,432)\,$MeV, $\sqrtsNN = 6\,$GeV for the lower bound. To our knowledge, this is the first time that an equilibrium 
baseline on the freeze-out line and the kurtosis at freeze-out is available, which reflects the phase structure and thermodynamics within 
direct computations from functional QCD.

However, we wish to emphasise again that our results only provide a first step towards the full picture. In order to compare oranges with oranges, we need to find a way to extract proton susceptibilities in our framework. This is work in progress. Furthermore, non-equilibrium effects at
the critical point, see e.g.~\cite{Stephanov:2011pb, Mukherjee:2015swa, Sieke:2024dns} may affect the shape of the kurtosis significantly even 
at a sizeable distance from the CEP. They need to be taken into account to further reduce the gap between theory and experiment. \\[-2ex]

%%%%%%%%%%%%%%%%%%%%%%%%%%%%
\textit{Acknowledgements.---} 
%%%%%%%%%%%%%%%%%%%%%%%%%%%%%%%%%
%
We thank Szabolcs Bors\'{a}nyi, Wei-jie Fu, Xiaofeng Luo, Nu Xu and the members of the fQCD collaboration \cite{fQCD} for discussions. This work is funded by the National Science Foundation of China under the grants No. 12305134,  12247107  and 12175007, the Deutsche Forschungsgemeinschaft (DFG, German Research Foundation) under Germany’s Excellence Strategy EXC 2181/1 - 390900948 (the Heidelberg STRUCTURES Excellence Cluster), the Collaborative Research Centre SFB 1225 - 273811115 (ISOQUANT). 
CSF has been supported by the Deut\-sche Forschungsgemeinschaft (DFG, German Research  Foundation) through the Collaborative Research Center TransRegio CRC-TR 211 `Strong-interaction matter under extreme conditions' and the individual grant FI 970/16-1.

\bibliography{refs}

\clearpage

%\appendix 
\renewcommand{\thesubsection}{{S.\arabic{subsection}}}
\setcounter{section}{0}
%\titleformat*{\section}{\centering \Large \bfseries}

\onecolumngrid

%\renewcommand{\thesubsection}{{S.\arabic{subsection}}}
%\setcounter{section}{0}
%\setcounter{equation}{0}

%\onecolumngrid

%%%%%%%%%%%%%%%%%%%%%%%%%%%%%%%%%%%
\section*{Supplemental Materials}

The Supplemental Material provides some details of our theoretical set-up. In \Cref{app:extraction}, we explain the details of extracting the freeze-out parameters from the cumulants measured in BES experiment together with the error estimates. In \Cref{app:baselines}, we summarise the predictions of baryon or proton number kurtosis from different studies on the market, and discuss the possible indications.

%%%%%%%%%%%%%%%%%%%%%%%%%%%%%%%%%%%
\subsection{Methodology and systematic error estimate for the systematic extraction of freeze-out temperature and chemical potential from the cumulants in BES experiments}
\label{app:extraction}

\begin{figure}[b]
  \centering
  \includegraphics[width=0.475\columnwidth]{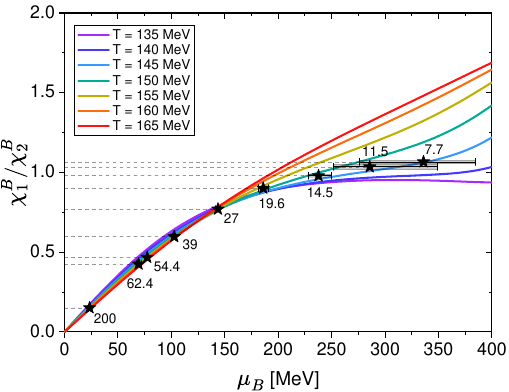}
  \includegraphics[width=0.475\columnwidth]{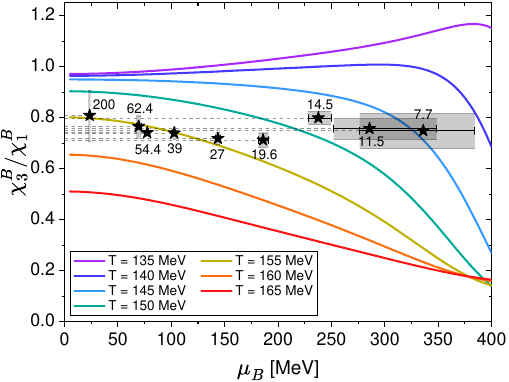}
  \caption{ Net-baryon number susceptibility ratios $\chi_1^B/\chi_2^B$ and $\chi_3^B/\chi_1^B$, compared with the experimental data of net-proton cumulant ratios $C_1/C_2$ and $C_3/C_1$ from BES-I $\sqrtsNN$ = 200 to 7.7 GeV, whose mean values are denoted by the dashed horizontal lines with the collision energies in GeV unit marked by the numbers alongside. The extracted freeze-out temperature $T_f$ and $\mu_{B,f}$ are denoted by stars, with the error bars denoted by the shaded rectangles. The vertical height of the rectangle corresponds to the experimental error at each $\sqrtsNN$. }\label{fig:cumulant-ratio-compare-bes1}
\end{figure}

Our extraction of the freeze-out parameters can be visualised in the $(T,\mu_B)$ plane. 
Specifically, the cumulant ratios $R_{12}^{B} = \chi_1^B/\chi_2^B$ and $R_{31}^{B} = \chi_3^B/\chi_1^B$ at a given temperature correspond to two types of isothermal curves respectively, see \Cref{fig:cumulant-ratio-compare-bes1}. 
As discussed in the main text, it is clear from \Cref{fig:cumulant-ratio-compare-bes1} that these two types of isotherms are sensitive to different thermodynamic parameters in the small $\mu_B$-region. Specifically, the $R_{12}^{B}$-curve depends almost exclusively on $\mu_B$, while the $R_{31}^{B}$ curve is more sensitive  more to temperature changes for $\mu_B \lesssim 150$\,MeV. Towards larger $\mu_B$, such a distinction is no longer possible, as both ratios show $T$- and $\mu_B$-dependences. 
Still, the two cumulant ratios offer a systematic way to extract the freeze-out temperature and chemical potential when contrasted with the respective experimental measurement. 

\begin{figure}[t]
  \centering
  \includegraphics[width=0.475\columnwidth]{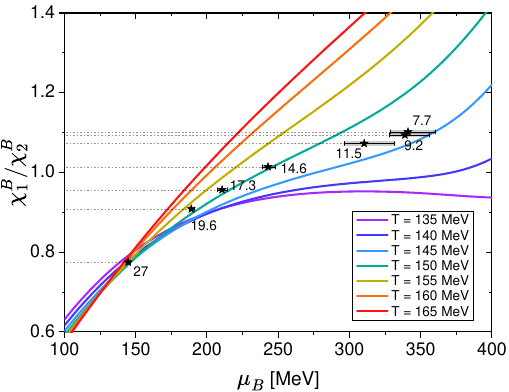}
  \includegraphics[width=0.475\columnwidth]{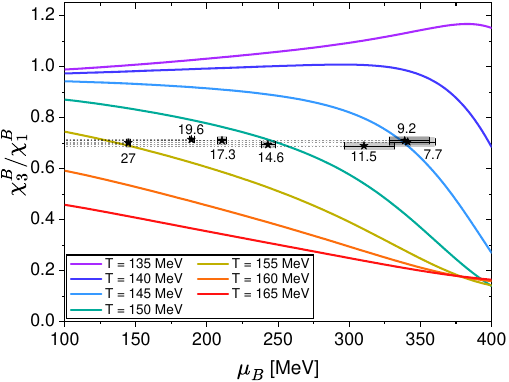}
  \caption{ Net-baryon number susceptibility ratios $\chi_1^B/\chi_2^B$ and $\chi_3^B/\chi_1^B$, compared with the experimental data of net-proton cumulant ratios $C_1/C_2$ and $C_3/C_1$ from BES-I $\sqrtsNN$ = 27 to 7.7 GeV. All symbols have the same meaning as those illustrated in \Cref{fig:cumulant-ratio-compare-bes1}. }\label{fig:cumulant-ratio-compare-bes2}
\end{figure}

In \Cref{fig:cumulant-ratio-compare-bes1} we show our results compared with the STAR BES-I data of the net-proton cumuant ratios $C_1/C_2$ and $C_3/C_1$ measured from $\sqrtsNN$ = 200 GeV to 7.7 GeV. The error bars of the experimental data are marked by the height of the shaded rectangles, which are the combined statistical and systematic errors with the root of their sum square~(RSS): 
\begin{align}
\sigma_{\textrm{tot.}} = \sqrt{ \sigma_{\textrm{stat.}}^2 + \sigma_{\textrm{sys.}}^2 }\,. \label{eq:total-error-exp}
\end{align}
Similar comparisons with the STAR BES-II data are provided in \Cref{fig:cumulant-ratio-compare-bes2} from $\sqrtsNN$ = 27 GeV to 7.7 GeV. In this regime, the experimental data have improved statistics and narrowed errors. 

\begin{figure}[t]
  \centering
  \includegraphics[width=0.47\columnwidth]{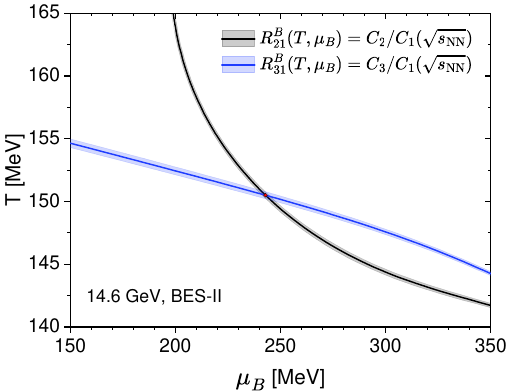}
  \includegraphics[width=0.47\columnwidth]{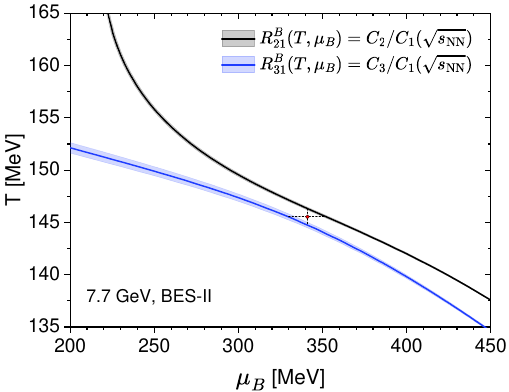}
  \caption{Illustration of the freeze-out extraction and its error estimate using the $T$-$\mu_B$ trajectories of $R_{21} = \chi_2^B/\chi_1^B$ and $R_{31}=\chi_3^B/\chi_1^B$ that match the BES-II cumulant data $C_2/C_1$ and $C_3/C_1$ at $\sqrtsNN$ = 14.6 GeV and 7.7 GeV. For 7.7 GeV the optimised $\mu_{B,f}$ is marked out as the dot where the $T_f$ difference between the two trajectories is minimised. }\label{fig:extract-error-demo}
\end{figure}

The systematic extraction of the freeze-out temperature and chemical potential is done as follows: For a fixed experimental cumulant ratio of $C_2/C_1$ or $C_3/C_1$, \labelcref{eq:freezeout-extract-C1C2} and \labelcref{eq:freezeout-extract-C3C1} define trajectories within the $(T,\mu_B)$ plane in QCD phase diagram. 
Then, the freeze-out condition $(T_f,\mu_{B,f})$ at each collision energy is determined by the intersection between the $C_2/C_1$ trajectory and the $C_3/C_1$ trajectory, which is illustrated in \Cref{fig:extract-error-demo}. 
For most $\sqrtsNN$, the two trajectories intersect each other, and the intersection point is defined as the mean value of the freeze-out $(T,\mu_B)$ in our extraction.
As an illustrative example, we show the situation for $\sqrtsNN = 14.6\,$GeV from BES-II data in the left panel of \Cref{fig:extract-error-demo}. 

We proceed with a discussion of impact of the experimental errors on the extraction. This error induces shifts of the trajectories around those determined by the mean cumulant ratios. For $C_2/C_1$ and $C_3/C_1$ the error is visualized as the gray band and blue band in \Cref{fig:extract-error-demo} respectively. 
The experimental error is accommodated by the overlap region of the gray band and blue band in the $(T,\mu_B)$ plane, i.e.~the overlapped region for the two trajectory bands in the left panel of \Cref{fig:extract-error-demo}. 
Correspondingly, the upper and lower bounds of this region are taken as the extraction errors of the freeze-out points. In  \Cref{fig:cumulant-ratio-compare-bes1} the error bars shown are combination of the experimental and extraction errors. 

However, going down to quite low collision energies, the $C_2/C_1$ trajectory and $C_3/C_1$ trajectory do not necessarily intersect with each other. This indicates a mismatch between the equilibrium QCD fluctuations and the experimental data, which is found for BES-I 7.7\,GeV and for BES-II 9.2 and 7.7\,GeV. We show the case of BES-II 7.7\,GeV in the right panel of \Cref{fig:extract-error-demo}, where a maximal gap is found between the $C_2/C_1$ and $C_3/C_1$ trajectories. 
To our understanding, this mismatch potentially comes from the difference between net-baryon and net-proton cumulants, as well as from the deviation of the final state after freeze-out from  thermodynamic equilibrium at such a low $\sqrtsNN$ in the experiment. 
Still, there exists an optimised $\mu_{B,f}$ where the gap between the two freeze-out temperatures $T_f$ from \labelcref{eq:freezeout-extract-C1C2,eq:freezeout-extract-C3C1} is minimised. For the lowest $\sqrtsNN$ = 7.7 GeV in BES-I and BES-II, the minimal gap is $\Delta T_f \approx 2$\,MeV. 
We consider the mean pair $(T_f,\mu_{B,f})$ from \labelcref{eq:freezeout-extract-C1C2,eq:freezeout-extract-C3C1} and the optimised $\mu_{B,f}$ as an estimate of the mean freeze-out temperature and chemical potential. The uncertainty in the $T$- and $\mu_B$-directions between the gap of the two trajectories defines the extraction error. This is visualised as the short-dashed lines in the right panel of \Cref{fig:extract-error-demo}.
The experimental error can be included as the error band of the two trajectories, which is summed up with the extraction error to define the total error for BES-I 7.7 GeV and BES-II 9.2 and 7.7 GeV.

%%%%%%%%%%%%%%%%%%%%%%%%%%%%%%%%%
\subsection{Baselines of net-baryon and (net-)proton kurtosis in the beam energy scan}
\label{app:baselines}

\begin{figure}[h]
  \centering
\includegraphics[width=0.47\columnwidth]{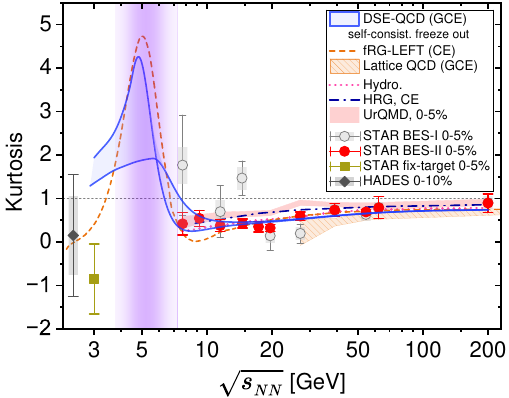}
  \caption{Collection of net-baryon and (net-)proton number kurtosis baseline predictions from $\sqrtsNN = $ 200\,GeV to 3\,GeV, compared with the experimental data from STAR 0-5\% central collisions in BES-I~\cite{STAR:2020tga}, BES-II~\cite{STAR:2025zdq}, STAR fixed-target ($-0.5<y<0$ rapidity window)~\cite{STAR:2021fge} and HADES (0-10\% centrality and $|y|<0.5$)~\cite{HADES:2020wpc}. The theoretical baselines include the results from the present work, QCD-assisted fRG-LEFT (canonical ensemble)~\cite{Fu:2023lcm}, lattice QCD~\cite{Bazavov:2020bjn}, HRG~\cite{Braun-Munzinger:2020jbk}, UrQMD~\cite{STAR:2021iop} and hydrodynamic simulations~\cite{Vovchenko:2021kxx}.}\label{fig:R42Freezeout-all}
\end{figure}

In this Supplement, we briefly summarise the baselines of net-baryon and (net-)proton number kurtosis obtained from previous studies in \Cref{fig:R42Freezeout-all}, which are also compared with the result in the present work. 
On the theoretical side, several baselines have been obtained from QCD computations for the net-baryon number kurtosis, including  functional approaches, the QCD-assisted fRG-low energy effective theory (LEFT) (canonical ensemble)~\cite{Fu:2021oaw,Fu:2023lcm}, and lattice QCD extrapolation results at relatively high $\sqrtsNN$~\cite{Bazavov:2020bjn}. 
We also show net-proton number kurtosis baselines from the hadron resonance gas (HRG) model~\cite{Braun-Munzinger:2020jbk}, UrQMD~\cite{STAR:2021iop} and hydrodynamic simulations~\cite{Vovchenko:2021kxx}. The experimental data includes the net-proton and proton number kurtosis from the STAR beam energy scan for Au+Au collisions (phase-II~\cite{STAR:2025zdq} and phase-I~\cite{STAR:2020tga}), as well as the proton number kurtosis from the STAR fixed target experiment at $3$\,GeV~\cite{STAR:2021fge} and from the HADES experiment at $2.4$\,GeV~\cite{HADES:2020wpc}. 

In the range of 3 to 7.7\,GeV, the results for the kurtosis may indicate potential new phases in the QCD phase structure. At present, the theoretical baseline predictions within this range are only available from the present work and from the QCD-assisted fRG-LEFT, and it is remarkable that both of them qualitatively agree on a peak structure of kurtosis and its location. 
Note that this is in line with the small difference between the upper bound of our freeze-out line and the fitted freeze-out line in~\cite{Andronic:2017pug} (which is the input of fRG-LEFT baseline), as indicated in \Cref{fig:freezeout-T-muB} in the main text. 

We envisage, that a combined analysis from different QCD computations in equilibrium and their systematic extension towards proton number cumulants, will finally lead to a robust quantitative constraint on the proton number kurtosis baseline in equilibrium. 
As discussed in the main text, this is a crucial first step before mapping out oranges (protons) with oranges (protons) between the quantitative theoretical description and precision measurements in experiment. Finally, of course, the theoretical studies have to be extended towards non-equilibrium QCD. 

\end{document}